\documentclass[journal]{IEEEtran}

\ifCLASSINFOpdf
\else
   \usepackage[dvips]{graphicx}
\fi
\usepackage{url}
\usepackage{cite}
\usepackage{amsmath,amssymb,amsfonts}
\usepackage{algorithmic}
\usepackage{graphicx}
\usepackage{textcomp}
\usepackage{xcolor}
\usepackage{graphicx}
\usepackage{multirow}
\usepackage{booktabs}
\usepackage{threeparttable}
\graphicspath{{figs/}}

\begin{document}

\title{Scale This, Not That: Investigating Key Dataset Attributes for Efficient Speech Enhancement Scaling}

\author{Leying Zhang, \IEEEmembership{Student Member, IEEE}, Wangyou Zhang, \IEEEmembership{Member, IEEE}, Chenda Li, \IEEEmembership{Member, IEEE} \\ and Yanmin Qian, \IEEEmembership{Senior Member, IEEE}
\thanks{Leying Zhang, Wangyou Zhang, Chenda Li, and Yanmin Qian are with Shanghai Jiao Tong University, Shanghai, 200240, China.  (e-mail: zhangleying,  wyz-97, lichenda1996, yanminqian@sjtu.edu.cn).
}}

\markboth{Journal of \LaTeX\ Class Files, Vol. 14, No. 8, August 2015}
{Shell \MakeLowercase{\textit{et al.}}: Bare Demo of IEEEtran.cls for IEEE Journals}
\maketitle

\begin{abstract}
Recent speech enhancement models have shown impressive performance gains by scaling up model complexity and training data. However, the impact of dataset variability (e.g. text, language, speaker, and noise) has been underexplored. Analyzing each attribute individually is often challenging, as multiple attributes are usually entangled in commonly used datasets, posing a significant obstacle in understanding the distinct contributions of each attribute to the model’s performance. To address this challenge, we propose a generation-training-evaluation framework that leverages zero-shot text-to-speech systems to investigate the impact of controlled attribute variations on speech enhancement performance. It enables us to synthesize training datasets in a scalable manner while carefully altering each attribute. Based on the proposed framework, we analyze the scaling effects of various dataset attributes on the performance of both discriminative and generative SE models. Extensive experiments on multi-domain corpora imply that acoustic attributes (e.g., speaker and noise) are much more important to current speech enhancement models than semantic attributes (e.g., language and text), offering new insights for future research.
\end{abstract}


\IEEEpeerreviewmaketitle

\section{Introduction}
\label{sec:intro}
As one of the fundamental components of speech processing systems, the task of speech enhancement (SE) has been extensively studied over the past decades. SE aims to remove undesired noisy signals from the input speech, thus improving the perceptual quality and intelligibility~\cite{loizou2007speech,zhang2023toward,zhang2024improving}. Recent SE models have demonstrated remarkable performance by enhancing the scalability in terms of architecture, model complexity, model size, and the training data size~\cite{zhang2024beyond,chen2024complexity,gonzalez2024effect}. As indicated in~\cite{zhang2024beyond}, the variability in the dataset remains an important issue when scaling up SE models, which received relatively less attention in previous research. 

Although existing studies have highlighted the importance of data size, they often overlooked the role of different dataset attributes (e.g., speaker identity, language, text), and it remains unclear which attributes should be prioritized when scaling up.
The efficiency of data scaling in SE research is thus limited, lacking well-established guidance of data selection, which can be addressed by analyzing the scaling effect of each individual dataset attribute.
However, the most commonly used SE corpora have largely entangled different attributes in the dataset, making it difficult to analyze each individual attribute without altering others. Despite the importance of the above issue, no effective framework has been proposed to address it.

On the other hand, zero-shot text-to-speech (TTS) aims to generate natural speech for unseen speakers with given speech prompts and text transcription~\cite{ju2024naturalspeech}. Recent advancements in deep learning have greatly improved the quality of TTS synthesized speech, approaching human-level naturalness ~\cite{wang2023neural,le2024voicebox,casanova2024xtts,anastassiou2024seed,lajszczak2024base}. Apart from speech synthesis, TTS models have also been widely used for augmenting the training data in various speech-related tasks such as automatic speech recognition (ASR)~\cite{fazel2021synthasr,eigenschink2023deep,rossenbach2024effect} and speaker verification~\cite{du2021synaug}.
Despite these successful applications, TTS models have rarely been used for data generation in SE research. This is primarily due to the data simulation nature of SE approaches, where the training data can be easily simulated by mixing clean speech and noise samples. Since these samples can be collected from either existing corpora or the Internet, it is often unnecessary to utilize TTS models to generate synthetic speech samples.
Another concern is that TTS models may introduce artifacts in synthetic speech samples, resulting in sub-optimal SE performance.
In this paper, however, we will show that the TTS model can be used to address the challenge in the preceding paragraph, enabling analysis of each individual dataset attribute in SE scaling without sacrificing much performance.

This paper aims to reveal the key dataset attributes with the largest impact on SE performance when scaling up the training data size. As mentioned earlier, addressing this problem can better facilitate data selection when scaling up, thus improving the efficiency and SE performance.
The contribution of this paper can be summarized as follows:

1) We proposed a three-stage analysis pipeline, which combines a zero-shot text-to-speech model and the speech enhancement model to enable analysis of the scaling effect of individual dataset attributes.

2) We verified the feasibility of training SE models based on purely synthetic speech data, which shows comparable performance to models trained on real-world speech data.

3) We investigated the scaling effects of four distinct attributes (i.e., text, language, speaker, and noise) on two representative SE models, revealing that they are largely text- and language-independent while being sensitive to speaker and noise diversity.

\section{Related Work}

\subsection{Zero-Shot Text-to-Speech System}

Zero-shot text-to-speech (ZS-TTS) technologies aim to generate human-like natural speech with voice characteristics prompted by the context. While most ZS-TTS models support a single language, there is growing interest in developing multi-lingual models. XTTS\footnote{https://github.com/coqui-ai/TTS}~\cite{casanova2024xtts} is an open-source, pre-trained multi-lingual TTS system that achieved state-of-the-art results among publicly available models across 16 languages, requiring only a 6s voice segment as the speaker prompt. 

\vspace{-0.3cm}
\subsection{Speech Enhancement System}

In the past decade, various deep learning techniques have been introduced to the SE task, with approaches generally classified into discriminative and generative methods~\cite{JET392}. Discriminative methods, such as Conv-TasNet~\cite{luo2019convtasnet} and band-split recurrent neural network (BSRNN)~\cite{luo2023bsrnn}, directly learn to predict the clean speech from noisy speech, achieving substantial advancements. In contrast, generative methods, particularly diffusion models~\cite{sgmse2023, zhang2024ddtse, lu2022conditional,li2024diffusion}, model the distribution of clean speech conditioned by noisy speech. Diverse speech can be generated by sampling from the learned distribution. In this study, we adopt BSRNN and SGMSE~\cite{sgmse2023} as representatives of the discriminative and generative approaches, respectively.

\vspace{-0.2cm}
\section{Method and Experimental Design}
To explore the effects of each dataset attribute, we apply a generation-training-evaluation framework. For a given attribute, we first generate the corresponding dataset and then train a specific model on this dataset. Finally, we evaluate both intrusive and non-intrusive metrics on multilingual datasets with both in-domain and out-of-domain noises.

\vspace{-0.4cm}
\subsection{Generation}

In order to analyze the impact of each attribute in the speech data, we generate a series of purely synthetic training datasets by utilizing a pre-trained ZS-TTS model (i.e., XTTS), which can precisely and independently control the textual content, spoken language of content, and speaker identity via prompting. In each generated dataset below, we keep all data attributes unaltered except for one, which can be any one of the aforementioned attributes\footnote{We will open-source the generation codes for future research: \url{https://reurl.cc/26bKGr}.}. All generated datasets have the same number of utterances (denoted as $m$) as $\mathcal{R}$.

Firstly, to explore the feasibility of purely synthetic data for SE training, we generate a synthetic dataset $\mathcal{F}$ which shares the same data attributes as $\mathcal{R}$, except that it is totally synthesized.

To manipulate the \textbf{textual content}, we generate multiple datasets $\mathcal{T}_n$ with $n$ unique transcriptions from $\mathcal{R}$. If $n < m$, we reuse the $n$ transcriptions multiple times to match the total number of utterances. We strictly constrain the total number of words between $\mathcal{R}$ and $\mathcal{T}_n$ to be the same to roughly retain the overall duration of the speech as in $\mathcal{R}$.

\label{sec:language-new}
To manipulate the \textbf{spoken language},  we generate multiple datasets $\mathcal{L}_p$ with $p$\,($1\leq p \leq 10$) languages, starting from a single language (i.e., English).
If $p > 1$, we manually select $p-1$ new languages from Chinese, Czech, German, Spanish, French, Italian, Polish, Russian, and Japanese, all supported by XTTS. 
We then synthesize $\frac{m(11-p)}{10}$ utterances for English and $\frac{m}{10}$ utterances for the newly selected $p-1$ languages, which together compose the final multilingual dataset.
Similarly, we also constrain the word count and speech duration in all datasets to be as close as possible.

To manipulate the \textbf{speaker diversity}, we generate multiple datasets $\mathcal{S}_s$ with randomly selected $s$ speakers.
Since multiple utterances from the same speaker can exist in the source dataset $\mathcal{R}$, we explore two modes for speaker-prompting-based generation: single-prompt and multi-prompt modes.
For the single-prompt mode, only one utterance of each speaker is chosen as the speaker prompt, which will be reused $\frac{m}{s}$ times. 
For the multi-prompt mode, $\frac{m}{s}$ unique utterances of each speaker are chosen as speech prompts, each used only once.

To analyze the \textbf{noise attribute}, we randomly sample noise subsets $\mathcal{J}_t$ from the whole noise dataset $\mathcal{N}$ given a desired total duration $t$. This allows for the exploration of SE performance with different noise durations. Alternatively, we sample multiple noise subsets $\mathcal{K}_{t,k}$ given a desired total duration $t$ and a given noisy types $k$, highlighting the impact of noise diversity w.r.t. noise types.

In practice, when analyzing speech attributes (e.g., text, language, and speaker), we simulate paired noisy-clean data before training following the fixed LibriMix simulation pipeline~\cite{cosentino2020librimix}, with the real speech dataset $\mathcal{R}$, the synthetic speech dataset $\mathcal{F}, \mathcal{T}, \mathcal{L}, \mathcal{S}$, and real environmental noises $\mathcal{N}$.
When analyzing noise attributes, however, we simulate paired data with the speech dataset and the selected noise dataset  $\mathcal{J}_t$ and $\mathcal{K}_{t,k}$ on the fly during training. We use LibriSpeech~\cite{panayotov2015librispeech} dataset train-100 set as the real speech $\mathcal{R}$. All noises are sampled from WHAM!~\cite{wichern2019wham} and TUT Urban Acoustic Scenes 2018~\cite{mesaros2016tut}  datasets. We generated multiple synthetic datasets using the XTTS model~\cite{casanova2024xtts}, employing speaker prompts extracted from LibriSpeech and transcriptions from LibriSpeech (English) and CommonVoice 11~\cite{ardila2019common} (other languages). Detailed dataset configuration is listed in Table \ref{tab:train-data}.

\begin{figure*}[htb]
  \centering
\centerline{\includegraphics[width=.90\linewidth]{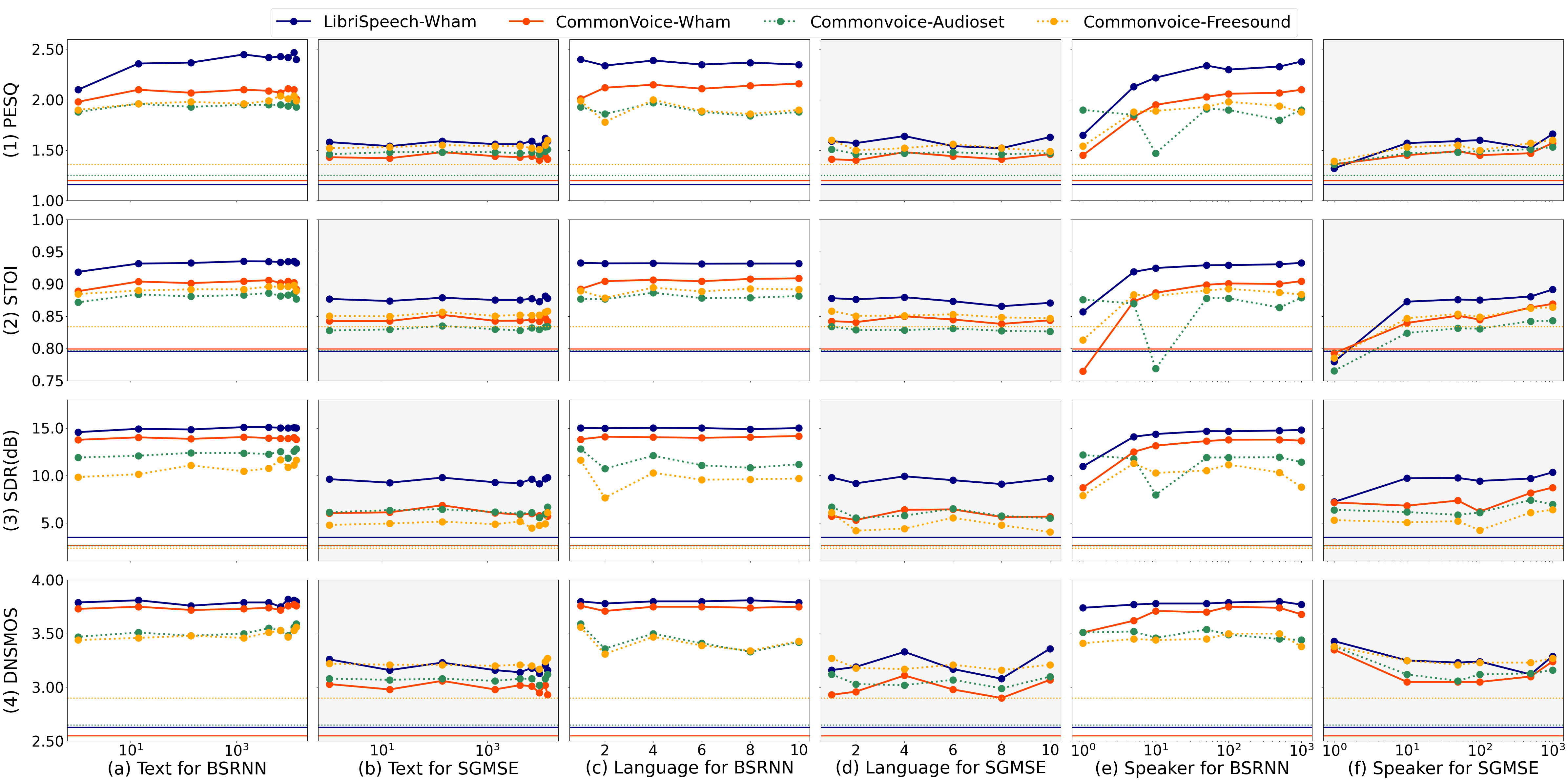}}
\vspace{-1em}
\caption{Analysis of data variability on text, language and speaker for both discriminative (BSRNN) and generative (SGMSE) speech enhancement models. The four horizontal lines in each plot represent the performances without any enhancement.}
\label{fig:main}
\vspace{-1em}
\end{figure*}

\begin{table}[t]
\footnotesize
    \centering
    \caption{Training Dataset Configuration. }
    \label{tab:train-data}
    \setlength\tabcolsep{1.5pt}
    \begin{tabular}{l|l|ccc|ccccc}
        \toprule
        \multirow{2}*{Speech} &  \multirow{2}*{Noise} & \multicolumn{3}{c|}{Source Dataset} &  \multicolumn{5}{c}{Attribute Configuration} \\
        ~ &~ & Spk & Text  & Noise & \#Text & \#Spk & \#Lang & \#Noise & \#NT \\ \midrule
        $\mathcal{R}$ & $\mathcal{N}$& LS & LS & W & 13900 & 251 & 1 & 58h & K  \\ 
        $\mathcal{F}$ & $\mathcal{N}$ & LS & LS & W & 13900 & 251 & 1 & 58h  & K\\ \midrule
        $\mathcal{T}_n$ & $\mathcal{N}$ & LS & LS & W & [1,13900] & 251 & 1 & 58h & K \\ 
         $\mathcal{S}_s$ & $\mathcal{N}$ & LS & LS & W & 13900 & [1,1000] & 1 & 58h & K \\ 
         $\mathcal{L}_p$ & $\mathcal{N}$ & LS & LS,CV & W & 13900 & 251 & [1,10] & 58h & K \\ 
         $\mathcal{F}$ & $\mathcal{J}_t$ & LS & LS & W,T& 13900 & 251 & 1 & [0.3h,82h] & K \\
         $\mathcal{F}$ & $\mathcal{K}_{t,k}$  & LS & LS & W,T& 13900 & 251 & 1 & [0.3h,82h] & [1,K] \\ \bottomrule
        \end{tabular}
        \begin{tablenotes}
        \footnotesize
        \item * We abbreviate the dataset of LibriSpeech, CommonVoice 11, WHAM!, TUT as LS, CV, W, T. We abbreviate \#N as the total number of Noise dataset (hours). We abbreviate \#NT as the number of noise types. 
        \end{tablenotes}
\end{table}

\vspace{-0.4cm}
\subsection{Training}
We selected two representative SE models for our investigation: BSRNN and SGMSE. 
We train all models for 40 epochs with the ESPnet Toolkit\footnote{\url{https://github.com/espnet/espnet}}~\cite{li2020espnet} with a batch size of 6 on the previously generated noisy-clean data pairs. We set the learning rate of BSRNN and SGMSE to 1e-3 and 1e-4, respectively. We utilize the Adam optimizer and a single Nvidia A10 GPU for training and evaluation. We fix the hyper-parameters for all experiments.


\vspace{-0.2cm}
\subsection{Evaluation}
To assess the generalizability and effectiveness, our evaluation data consists of real-world multilingual speech data with both in- and out-of-domain noises. Specifically, we employ four evaluation scenarios. The LibriMix official test set is for in-domain evaluation. We additionally simulate three out-of-domain multilingual evaluation datasets based on CommonVoice 11 speech data and noises from WHAM!, Freesound~\cite{fonseca2017freesound}, and Audioset~\cite{gemmeke2017audio}, where the signal-to-noise ratio (SNR) ranges from -5 to 10 dB.
The multilingual evaluation datasets cover a total of 10 languages mentioned in Section \ref{sec:language-new}.
We evaluate with four metrics: perceptual evaluation of speech quality (PESQ)~\cite{rix2001perceptual}, short-time objective intelligibility (STOI) \cite{jensen2016algorithm}, signal-to-distortion ratio (SDR)~\cite{Performance-Vincent2006} and DNSMOS~\cite{reddy2021dnsmos}. All metrics are the higher, the better.

\section{Results and Analysis}
\subsection{Validity of Purely Synthetic Data for SE Training}

In the literature, SE models are generally trained on real-world speech data. However, purely synthetic speech data play a vital role in our investigation.
Thus, in this subsection, we first compare the performance of models trained on real-world speech data $\mathcal{R}$ with those trained solely on synthetic speech data $\mathcal{F}$. Table \ref{tab:validity} indicates that for both discriminative models (i.e., BSRNN) and generative models (i.e., SGMSE), the models trained on purely synthetic speech data only exhibit a slight performance decline compared to those trained on real-world speech data.
Therefore, we can conclude that purely synthetic speech data based on a well-trained TTS model (e.g., XTTS) can be used for SE training without greatly degrading the performance.
This finding thus validates our later investigations based on purely synthetic speech data.

\begin{table}[t]
    \centering
    \caption{Performance Comparison of Real ($\mathcal{R}$) and Purely Synthetic data ($\mathcal{F}$) across different models on LibriMix test set}
    \label{tab:validity}
    \begin{tabular}{c|c|cccc}
    \toprule
        Model & Data & PESQ & STOI & SDR(dB) & DNSMOS  \\ \midrule
        \multirow{2}*{BSRNN}  & $\mathcal{R}$ & 2.76 & 0.95 & 15.93 & 3.79  \\ 
        ~ & $\mathcal{F}$ & 2.40 & 0.93 & 15.01 & 3.80  \\ \midrule
\multirow{2}*{SGMSE}  & $\mathcal{R}$ & 1.68 & 0.89 & 10.13 & 3.35  \\ 
        ~ & $\mathcal{F}$ & 1.59 & 0.88 & 9.80 &  3.16 \\ \bottomrule
    \end{tabular}
\end{table}

\vspace{-0.2cm}
\subsection{Effects of Text Variability}

As the spoken content in speech, text is highly associated with phoneme information, which plays an important role in various speech tasks (e.g., ASR, TTS).
It is natural to ask whether text also has a great impact on speech enhancement.
In this subsection, we explore the relationship between text diversity and the final SE performance.
Figure \ref{fig:main} (a, b) illustrates the performance of the SE models when trained with different numbers of unique transcriptions.  
Interestingly, our findings suggest that text content has a relatively minor influence on the performance of both discriminative and generative SE models. Reducing text diversity does not significantly affect SE performance. Even in an extreme case, where all utterances were generated using a single sentence as the transcription, the models' performances only show a slight decline for BSRNN and almost no decrease for SGMSE.

\vspace{-0.3cm}
\subsection{Effects of Language Variability}

The spoken language is another factor highly associated with phoneme information, where different languages may cover a different set of phonemes.
Figure \ref{fig:main} (c, d) illustrates the performance of the SE model trained with different numbers of languages while maintaining the total speech duration. We observe that both SE models can achieve decent overall performance in multilingual scenarios (covering ten languages) even if the model is trained only in English. Moreover, their performance remains relatively stable with the addition of new languages. This outcome is consistent with our findings on text variability, suggesting that semantic information may have a limited impact on the effectiveness of SE models. However, the above observation does not necessarily indicate the unimportance of multilingual data for SE training, as our evaluation data may not cover the entire phoneme set of all tested languages. Instead, we may conclude that it is safe to scale the SE training data by introducing new languages.

\begin{figure}[t]
  \centering
\centerline{\includegraphics[width=1.0\linewidth]{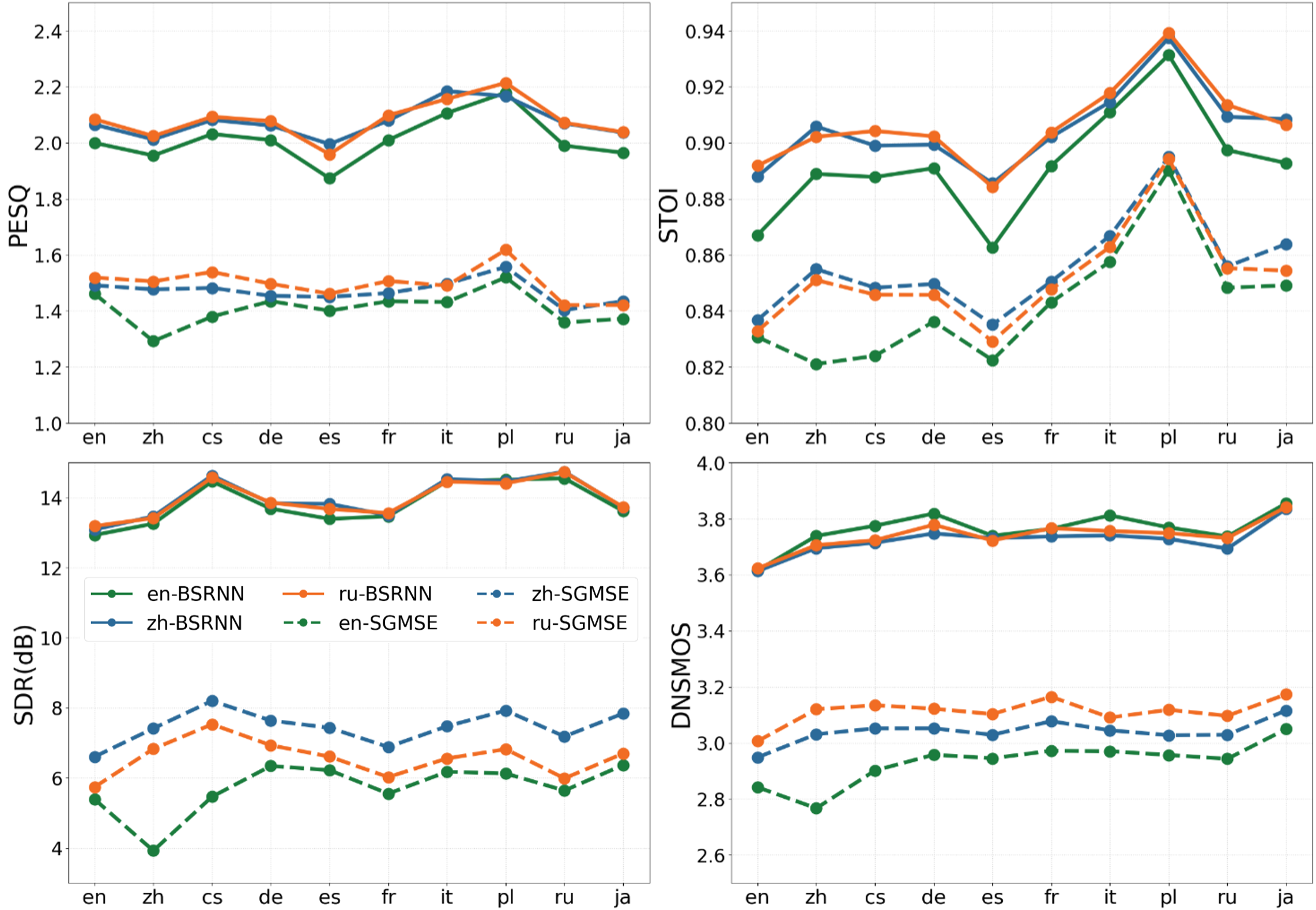}}
\vspace{-0.7em}
  \caption{Evaluaion on 10 different languages across models trained with English, Chinese and Russian}
\label{fig:language}
\end{figure}

Additionally, we train SE models exclusively on English, Chinese, and Russian data. We then evaluate their performance across ten different languages, as shown in Figure \ref{fig:language}. In contrast to previous studies~\cite{li2011comparative} on non-deep learning approaches, which show that systems trained in English do not transfer well to other languages, our results indicate that modern deep learning methods can generalize effectively to unseen languages. However, the performance of the generative model varies based on the training language. For instance, SGMSE trained on Chinese and Russian exhibits strong performance in English, while SGMSE trained on English shows poor generalization to Chinese. Therefore, a more detailed exploration with large-scale multilingual data is needed in future work.

\vspace{-1em}
\subsection{Effects of Speaker Variability}
\begin{figure}[t]
  \centering
\centerline{\includegraphics[width=1.0\linewidth]{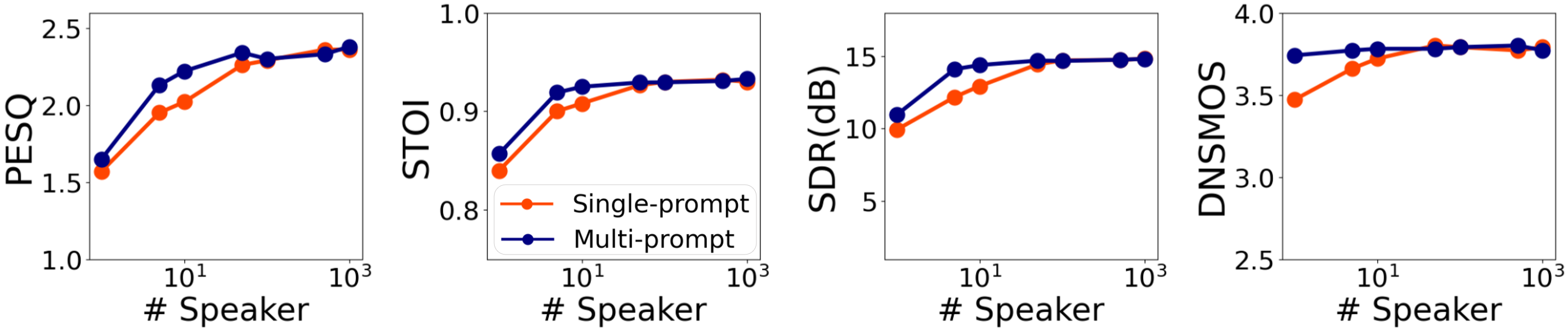}}
\vspace{-0.7em}
\caption{Analysis of prompt variability for given fixed number of speakers}
\label{fig:spk}
\end{figure}

\vspace{-0.3em}
Figure \ref{fig:main} (e, f) shows that the discriminative model, BSRNN, performs better with an increased number of speakers.
This observation highlights the importance of speaker diversity for SE training, which can provide rich acoustic variations for better generalization.
However, its performance saturates beyond 100 speakers due to the limitations of the total data size and model capacity. The generative model, SGMSE, follows a similar trend, though the effect is less pronounced.
Furthermore, in out-of-domain evaluation sets with unseen noises (Commonvoice-Audioset), the generalization capabilities of BSRNN and SGMSE do not show consistent improvement across all metrics except for STOI. This suggests that the richness of the speech signal offers limited benefits to improve model generalization in unseen noise conditions. Additionally, we assess the impact of intra-speaker variability in synthetic data generation. As shown in Figure \ref{fig:spk}, even with a fixed number of speakers, increasing the variety of prompts provided by each speaker leads to richer timbre diversity, resulting in improved speech enhancement performance, especially when few speaker's utterances are accessible.

\vspace{-1em}
\subsection{Effects of Noise Variability}

\begin{figure}[t]
  \centering
\centerline{\includegraphics[width=1.0\linewidth]{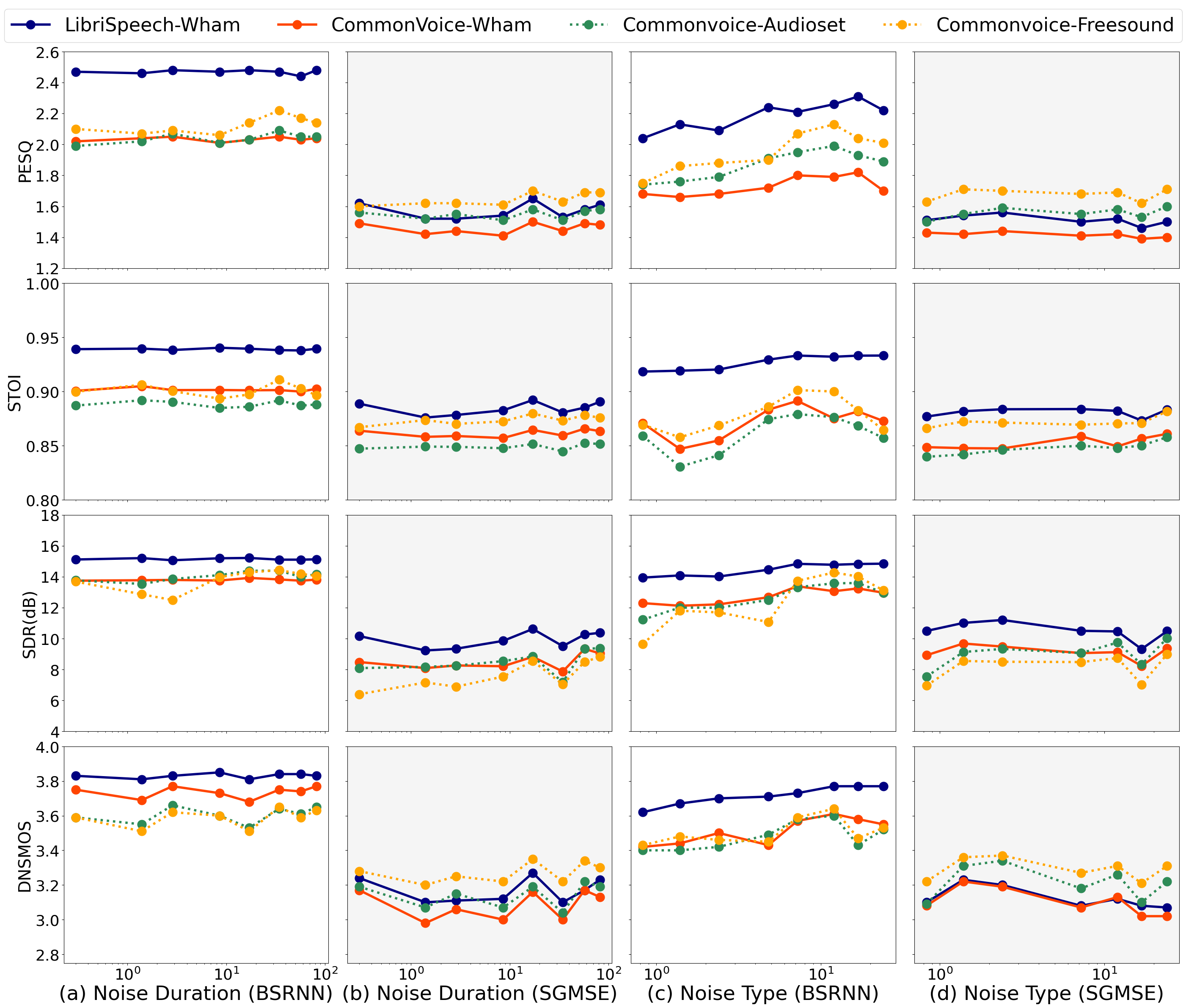}}
\vspace{-0.7em}
\caption{Analysis of effects of the noise type and noise duration}
\label{fig:noise}
\end{figure}

Finally, we investigate the relationship between noise diversity and SE performance.
We first examine how increasing the total duration of noise data while keeping the noise types constant affects model performance. As shown in Figure \ref{fig:noise} (a, b), the results on both in-domain and out-of-domain data remain stable, suggesting that merely adding more noise data, without enhancing its diversity, has a limited impact on the overall performance. In contrast, Figure \ref{fig:noise} (c, d) shows that expanding both the quantity and variety of noise types leads to improved performance for BSRNN, particularly on unseen noise. SGMSE shows less pronounced gains from the increased noise variety. These findings indicate that for training SE models, especially discriminative ones like BSRNN, enlarging noise types is more effective than only increasing the volume of noise data.

\vspace{-0.75em}
\section{Conclusion and Future work}
\vspace{-0.1em}
In this study, we proposed a generation-training-evaluation framework to explore the effects of various dataset attributes on the SE task. Our experiments demonstrated that text and language variabilities have a relatively limited influence on SE models, with performance remaining robust even under constrained linguistic diversity. Speaker variability improves SE performance, particularly with an increased number of speakers and timbre diversity. Noise variability, particularly through diverse noise types, proves essential for BSRNN’s generalization to unseen noise conditions, whereas SGMSE shows less sensitivity to noise variety. For future work, we plan to extend the analysis pipeline to a broader range of tasks and a larger data scale.

\bibliographystyle{IEEEtran}
\bibliography{refs}

\begin{thebibliography}{10}
\providecommand{\url}[1]{#1}
\csname url@samestyle\endcsname
\providecommand{\newblock}{\relax}
\providecommand{\bibinfo}[2]{#2}
\providecommand{\BIBentrySTDinterwordspacing}{\spaceskip=0pt\relax}
\providecommand{\BIBentryALTinterwordstretchfactor}{4}
\providecommand{\BIBentryALTinterwordspacing}{\spaceskip=\fontdimen2\font plus
\BIBentryALTinterwordstretchfactor\fontdimen3\font minus \fontdimen4\font\relax}
\providecommand{\BIBforeignlanguage}[2]{{%
\expandafter\ifx\csname l@#1\endcsname\relax
\typeout{** WARNING: IEEEtran.bst: No hyphenation pattern has been}%
\typeout{** loaded for the language `#1'. Using the pattern for}%
\typeout{** the default language instead.}%
\else
\language=\csname l@#1\endcsname
\fi
#2}}
\providecommand{\BIBdecl}{\relax}
\BIBdecl

\bibitem{loizou2007speech}
P.~C. Loizou, \emph{Speech enhancement: theory and practice}.\hskip 1em plus 0.5em minus 0.4em\relax CRC press, 2007.

\bibitem{zhang2023toward}
W.~Zhang, K.~Saijo, Z.-Q. Wang, S.~Watanabe, and Y.~Qian, ``Toward universal speech enhancement for diverse input conditions,'' in \emph{Proc. ASRU}, 2023, pp. 1--6.

\bibitem{zhang2024improving}
W.~Zhang, J.-w. Jung, and Y.~Qian, ``Improving design of input condition invariant speech enhancement,'' in \emph{Proc. ICASSP}, 2024, pp. 10\,696--10\,700.

\bibitem{zhang2024beyond}
W.~Zhang, K.~Saijo, J.-w. Jung, C.~Li, S.~Watanabe, and Y.~Qian, ``Beyond performance plateaus: A comprehensive study on scalability in speech enhancement,'' in \emph{Proc. Interspeech}, 2024, pp. 1740--1744.

\bibitem{chen2024complexity}
H.~Chen, J.~Yu, and C.~Weng, ``Complexity scaling for speech denoising,'' in \emph{Proc. ICASSP}, 2024, pp. 12\,276--12\,280.

\bibitem{gonzalez2024effect}
P.~Gonzalez, Z.-H. Tan, J.~{\O}stergaard, J.~Jensen, T.~S. Alstr{\o}m, and T.~May, ``The effect of training dataset size on discriminative and diffusion-based speech enhancement systems,'' \emph{IEEE Signal Processing Letters}, vol.~31, pp. 2225--2229, 2024.

\bibitem{ju2024naturalspeech}
Z.~Ju, Y.~Wang, K.~Shen, X.~Tan, D.~Xin, D.~Yang, Y.~Liu, Y.~Leng, K.~Song, S.~Tang \emph{et~al.}, ``{NaturalSpeech} 3: Zero-shot speech synthesis with factorized codec and diffusion models,'' in \emph{Proc. ICML}, 2024.

\bibitem{wang2023neural}
C.~Wang, S.~Chen, Y.~Wu, Z.~Zhang, L.~Zhou, S.~Liu, Z.~Chen, Y.~Liu, H.~Wang, J.~Li \emph{et~al.}, ``Neural codec language models are zero-shot text to speech synthesizers,'' \emph{arXiv preprint arXiv:2301.02111}, 2023.

\bibitem{le2024voicebox}
M.~Le, A.~Vyas, B.~Shi, B.~Karrer, L.~Sari, R.~Moritz, M.~Williamson, V.~Manohar, Y.~Adi, J.~Mahadeokar, and W.-N. Hsu, ``Voicebox: Text-guided multilingual universal speech generation at scale,'' in \emph{Advances in Neural Information Processing Systems}, vol.~36, 2024, pp. 14\,005--14\,034.

\bibitem{casanova2024xtts}
E.~Casanova, K.~Davis, E.~Gölge, G.~Göknar, I.~Gulea, L.~Hart, A.~Aljafari, J.~Meyer, R.~Morais, S.~Olayemi, and J.~Weber, ``{XTTS}: A massively multilingual zero-shot text-to-speech model,'' in \emph{Proc. Interspeech}, 2024, pp. 4978--4982.

\bibitem{anastassiou2024seed}
P.~Anastassiou, J.~Chen, J.~Chen, Y.~Chen, Z.~Chen, Z.~Chen, J.~Cong, L.~Deng, C.~Ding, L.~Gao \emph{et~al.}, ``Seed-{TTS}: A family of high-quality versatile speech generation models,'' \emph{arXiv preprint arXiv:2406.02430}, 2024.

\bibitem{lajszczak2024base}
M.~{\L}ajszczak, G.~C{\'a}mbara, Y.~Li, F.~Beyhan, A.~van Korlaar, F.~Yang, A.~Joly, {\'A}.~Mart{\'\i}n-Cortinas, A.~Abbas, A.~Michalski \emph{et~al.}, ``{BASE} {TTS}: Lessons from building a billion-parameter text-to-speech model on 100k hours of data,'' \emph{arXiv preprint arXiv:2402.08093}, 2024.

\bibitem{fazel2021synthasr}
A.~Fazel, W.~Yang, Y.~Liu, R.~Barra-Chicote, Y.~Meng, R.~Maas, and J.~Droppo, ``{SynthASR}: Unlocking synthetic data for speech recognition,'' in \emph{Proc. Interspeech}, 2021, pp. 896--900.

\bibitem{eigenschink2023deep}
P.~Eigenschink, T.~Reutterer, S.~Vamosi, R.~Vamosi, C.~Sun, and K.~Kalcher, ``Deep generative models for synthetic data: A survey,'' \emph{IEEE Access}, vol.~11, pp. 47\,304--47\,320, 2023.

\bibitem{rossenbach2024effect}
N.~Rossenbach, B.~Hilmes, and R.~Schl{\"u}ter, ``On the effect of purely synthetic training data for different automatic speech recognition architectures,'' \emph{arXiv preprint arXiv:2407.17997}, 2024.

\bibitem{du2021synaug}
C.~Du, B.~Han, S.~Wang, Y.~Qian, and K.~Yu, ``{SynAug}: Synthesis-based data augmentation for text-dependent speaker verification,'' in \emph{Proc. ICASSP}, 2021, pp. 5844--5848.

\bibitem{JET392}
A.~Yuliani, M.~F. Amri, E.~Suryawati, A.~Ramdan, and H.~Pardede, ``Speech enhancement using deep learning methods: A review,'' \emph{Jurnal Elektronika dan Telekomunikasi}, vol.~21, no.~1, pp. 19--26, 2021.

\bibitem{luo2019convtasnet}
Y.~Luo and N.~Mesgarani, ``Conv-{TasNet}: Surpassing ideal time--frequency magnitude masking for speech separation,'' \emph{IEEE/ACM transactions on audio, speech, and language processing}, vol.~27, no.~8, pp. 1256--1266, 2019.

\bibitem{luo2023bsrnn}
Y.~Luo and J.~Yu, ``Music source separation with band-split {RNN},'' \emph{IEEE/ACM Transactions on Audio, Speech, and Language Processing}, vol.~31, pp. 1893--1901, 2023.

\bibitem{sgmse2023}
J.~Richter, S.~Welker, J.-M. Lemercier, B.~Lay, and T.~Gerkmann, ``Speech enhancement and dereverberation with diffusion-based generative models,'' \emph{IEEE/ACM Transactions on Audio, Speech, and Language Processing}, vol.~31, pp. 2351--2364, 2023.

\bibitem{zhang2024ddtse}
L.~Zhang, L.~Y. Yao~Qian, H.~Wang, H.~Yang, S.~Liu, L.~Zhou, and Y.~Qian, ``{DDTSE}: Discriminative diffusion model for target speech extraction,'' \emph{Proc. SLT}, 2024.

\bibitem{lu2022conditional}
Y.-J. Lu, Z.-Q. Wang, S.~Watanabe, A.~Richard, C.~Yu, and Y.~Tsao, ``Conditional diffusion probabilistic model for speech enhancement,'' in \emph{Proc. ICASSP}, 2022, pp. 7402--7406.

\bibitem{li2024diffusion}
C.~Li, S.~Cornell, S.~Watanabe, and Y.~Qian, ``Diffusion-based generative modeling with discriminative guidance for streamable speech enhancement,'' \emph{arXiv preprint arXiv:2406.13471}, 2024.

\bibitem{cosentino2020librimix}
J.~Cosentino, M.~Pariente, S.~Cornell, A.~Deleforge, and E.~Vincent, ``Librimix: An open-source dataset for generalizable speech separation,'' \emph{arXiv preprint arXiv:2005.11262}, 2020.

\bibitem{panayotov2015librispeech}
V.~Panayotov, G.~Chen, D.~Povey, and S.~Khudanpur, ``Librispeech: an {ASR} corpus based on public domain audio books,'' in \emph{Proc. ICASSP}, 2015, pp. 5206--5210.

\bibitem{wichern2019wham}
G.~Wichern, J.~Antognini, M.~Flynn, L.~R. Zhu, E.~McQuinn, D.~Crow, E.~Manilow, and J.~Le~Roux, ``{WHAM!}: Extending speech separation to noisy environments,'' in \emph{Proc. Interspeech}, 2019, pp. 1368--1372.

\bibitem{mesaros2016tut}
A.~Mesaros, T.~Heittola, and T.~Virtanen, ``{TUT} database for acoustic scene classification and sound event detection,'' in \emph{24th European Signal Processing Conference (EUSIPCO)}, 2016, pp. 1128--1132.

\bibitem{ardila2019common}
R.~Ardila, M.~Branson, K.~Davis, M.~Kohler, J.~Meyer, M.~Henretty, R.~Morais, L.~Saunders, F.~Tyers, and G.~Weber, ``Common voice: A massively-multilingual speech corpus,'' in \emph{Proceedings of the 12th Language Resources and Evaluation Conference}, 2020, pp. 4218--4222.

\bibitem{li2020espnet}
C.~Li, J.~Shi, W.~Zhang, A.~S. Subramanian, X.~Chang, N.~Kamo, M.~Hira, T.~Hayashi, C.~Boeddeker, Z.~Chen, and S.~Watanabe, ``{ESPnet-SE}: End-to-end speech enhancement and separation toolkit designed for {ASR} integration,'' in \emph{Proc. SLT}, 2021, pp. 785--792.

\bibitem{fonseca2017freesound}
E.~Fonseca, J.~Pons~Puig, X.~Favory, F.~Font~Corbera, D.~Bogdanov, A.~Ferraro, S.~Oramas, A.~Porter, and X.~Serra, ``Freesound datasets: A platform for the creation of open audio datasets,'' in \emph{Proc. ISMIR}, 2017, pp. 486--493.

\bibitem{gemmeke2017audio}
J.~F. Gemmeke, D.~P. Ellis, D.~Freedman, A.~Jansen, W.~Lawrence, R.~C. Moore, M.~Plakal, and M.~Ritter, ``Audio set: An ontology and human-labeled dataset for audio events,'' in \emph{Proc. ICASSP}, 2017, pp. 776--780.

\bibitem{rix2001perceptual}
A.~W. Rix, J.~G. Beerends, M.~P. Hollier, and A.~P. Hekstra, ``Perceptual evaluation of speech quality (pesq)-a new method for speech quality assessment of telephone networks and codecs,'' in \emph{Proc. ICASSP}, vol.~2, 2001, pp. 749--752.

\bibitem{jensen2016algorithm}
J.~Jensen and C.~H. Taal, ``An algorithm for predicting the intelligibility of speech masked by modulated noise maskers,'' \emph{IEEE/ACM Transactions on Audio, Speech, and Language Processing}, vol.~24, no.~11, pp. 2009--2022, 2016.

\bibitem{Performance-Vincent2006}
E.~Vincent, R.~Gribonval, and C.~F{\'e}votte, ``Performance measurement in blind audio source separation,'' \emph{IEEE Transactions on Audio, Speech, and Language Processing}, vol.~14, no.~4, pp. 1462--1469, 2006.

\bibitem{reddy2021dnsmos}
C.~K. Reddy, V.~Gopal, and R.~Cutler, ``{DNSMOS}: A non-intrusive perceptual objective speech quality metric to evaluate noise suppressors,'' in \emph{Proc. ICASSP}, 2021, pp. 6493--6497.

\bibitem{li2011comparative}
J.~Li, L.~Yang, J.~Zhang, Y.~Yan, Y.~Hu, M.~Akagi, and P.~C. Loizou, ``Comparative intelligibility investigation of single-channel noise-reduction algorithms for {Chinese}, {Japanese}, and {English},'' \emph{The Journal of the Acoustical Society of America}, vol. 129, no.~5, pp. 3291--3301, 2011.

\end{thebibliography}

\end{document}